\documentstyle[12pt]{article}
\newcommand{\beq}{\begin{equation}}
\newcommand{\eeq}{\end{equation}}
\newcommand{\beqa}{\begin{eqnarray}}
\newcommand{\eeqa}{\end{eqnarray}}
\newcommand{\beqar}{\begin{eqnarray*}}
\newcommand{\eeqar}{\end{eqnarray*}}
\newcommand{\tr}{{\rm tr}}

\def \vr {\hat\varrho}
\def \avr {\varrho}

\def \la {\langle}
\def \ra {\rangle}
\def \up {\uparrow}
\def \down {\downarrow}
\def \l {{\cal L}}
\def \o {{\cal O}}
\def \g {{\cal G}}
\def \h {{\cal H}}
\def \s {{\cal S}}

\def \a {\alpha}
\def \b {\beta}

\def \d {\delta}

\begin{document}
\begin{titlepage}
\vspace{.5in}
\thispagestyle{empty}
\begin{flushright}
TP-95-004\\
quant-ph/9504019
\end{flushright}
\vspace{0.5in}
\begin{center}
{\bf\Large Unitary Evolution Between Pure and Mixed States}\\
\vspace{.4in}
B. Reznik
\footnote{\it e-mail: reznik@physics.ubc.ca}\\
\medskip
{\small\it  Department of Physics}\\
{\small \it University of British Columbia}\\
{\small\it 6224 Agricultural Rd. Vancouver, B.C., Canada
V6T1Z1}\\
\end{center}
\vspace{.5in}
\begin{center}
\begin{minipage}{5in}
\begin{center}
{\large\bf Abstract}
\end{center}
{\small
We propose an extended  quantum  mechanical formalism that is
based on
a wave operator $\vr$, which is related to the ordinary density
matrix  via  $\rho=\vr\vr^\dagger$.
This formalism allows a (generalized) unitary evolution between
pure and mixed states.
It also preserves  much of the
connection between symmetries and conservation laws.
The new formalism is illustrated for the case
of a two level system.
}
\end{minipage}
\end{center}
\end{titlepage}


Several proposals motivated by various considerations
for generalizing the  quantum mechanical formalism
have been made todate.
In these programs one disposes
of a fundamental quantum mechanical principle
such as linearity, locality, or unitarity.
Weinberg suggested a nonlinear generalization and
proposed   precision tests of nonlinear corrections to quantum
mechanics \cite{Weinberg}.
Motivated by the apparent breakdown of unitarity in the
black-hole
evaporation process, Hawking proposed that a synthesis of quantum
mechanics and general relativity requires giving up
unitarity
\cite{Hawking}, and to some extent locality \cite{Hawking2}.
A model which  gives up both properties,
was constructed by Marinov
\cite{Marinov}.
As a linear and  local phenomenological implementation of
Hawking's proposal, Ellis, Hagelin, Nanopoulos and Srednicki
(EHNS)
\cite{Ellis}, and  Banks, Peskin and Susskind (BPS) \cite{BPS},
suggested a  modified Liouville equation for the
density matrix
$\rho$.

In particular, BPS showed that the requirements of
linearity, locality in time,
and conservation  of probabilities, lead to
a modified equation with a ``generic form'':
\beq
i\hbar\partial_t\rho = [H, \rho] +
i\sum_{n,m}h_{nm}(Q_mQ_n\rho + \rho Q_mQ_n - 2Q_n\rho Q_m
),
\label{BPS}
\eeq
Here,  $Q_n$ are any Hermitian operators and $h_{nm}$ is
c-number  hermitian matrix.
A sufficient but not  necessary  condition ensuring the
positivity of $\rho$, is that the matrix $h$ is positive.
Equation (\ref{BPS}) does not preserve $\tr\rho^2$.
Thus pure  states can indeed evolve to mixed states  \cite{SUW}.

Similar equations  can be obtained  from ordinary quantum mechanics
for a  subsystem interacting with an environment \cite{sen}.
Nevertheless, when gravity is involved one can argue that the
relevant ``micro-environment'' is hidden by  black-hole
horizons    and is {\it in principle}  unobservable. This would render
equation (\ref{BPS}) a fundamental modification of quantum
mechanics,  rather then an artifact of interacting with an environment.

Modified evolutions  like (\ref{BPS}) were  applied in various
cases. EHNS proposed that the corrections induced might be observed in
the ultra-sensitive $K_0-\bar K_0$ system.
Furthermore, Ellis et. al.  \cite{Ellis2} and
Huet and Peskin \cite{Huet}
examined the possibility that the
observed CP violation in the $K_0-\bar K_0$ system
is, to some extent,  due to  non-quantum mechanical
corrections.
Related  modifications where also proposed in connection with the
``measurement problem'', in order to  generate a
von-Neumann reduction for macroscopic systems
\cite{GRW,Ellis3}.

In what follows,  we propose a different  approach.
It is also based on the Liouville equation but not for the
ordinary
density matrix.
It constitutes a  linear, local,
and unitary extension of  quantum mechanics.
To this end,  consider the density matrix in
ordinary quantum mechanics and focus first on the case of a pure
state.
By analogy with the relation $\rho=|\psi\ra\la\psi|$,
let us define the operator $\vr$  by \cite{symm}:
\beq
\rho = \vr\vr^\dagger.
\label{squar}
\eeq
If $\vr$ satisfies  a Liouville equation
\beq
i\hbar\partial_t \vr = [H, \vr],
\label{Liouville}
\eeq
it is easy to see that the density matrix $\rho = \vr\vr^\dagger$
also satisfies a  Liouville equation with the same Hamiltonian.
The initial condition may be specified in terms of the
``square root operator''
$\vr$, rather than $\rho$. Thus, if
the system is determined at $t=t_0$ by an ordinary  complete set
of measurements to be
in the state $|\psi_0\ra$, (or $\rho=|\psi_0\ra\la \psi_0|$),
 this sets the initial condition
for equation (\ref{Liouville}):
\beq
\vr(t=t_0) = |\psi_0\ra\la \psi_0|.
\label{initial}
\eeq
Now we observe that (\ref{squar}) and (\ref{initial})
imply $\rho(t=t_0)=\vr(t=t_0)$, and since both quantities
obey the  same equation of motion this relation holds at any
subsequent time. The expectation values of any observable $A$ is
obtained  by the standard expression:
\beq
\la A\ra  = {\tr A\rho \over \tr\rho} =
                 {\tr A\vr\over\tr\vr}.
\eeq
Hence eqs. (2-5) are equivalent to ordinary quantum mechanics
\cite{note}.

It is therefore interesting to question whether eq. (3) can now
be used  as a new starting point for a quantum mechanical extension.
We shall assume that  $\vr$ is from now on a
general operator (not necessarily a projector)
still obeying the initial condition (4), and that
expectation values are still obtained by the standard expression
\beq
\la A \ra = {\tr A\rho\over \tr\rho}.
\label{prob}
\eeq
The density matrix however
is from now on obtained via $\rho=\vr\vr^\dagger$.

The hermiticity and positivity of $\rho$ is automatically
ensured by $\rho=\vr\vr^\dagger$.
We need that the modified equation
conserve probabilities, i.e,
$\partial_t\tr\rho=\partial_t\tr\vr\vr^\dagger=0$, but
not necessarily purity.
The most general  linear \cite{note2} and  local generalization
of eq. (\ref{Liouville})  which satisfies this  condition
can be written as:
\beq
i\hbar\partial_t \vr = [H , \vr] + L\vr + \vr R + g_{ij} K_i\vr
K'_j.
\label{gLiouville}
\eeq
Here, $L$, $R$, $K_i$ and $K'_j$ are  any Hermitian operators,
$g_{ij}$ are real coefficients, and the summation convention
was used.

Eq.  (\ref{gLiouville})
implies that the density matrix  obeys:
\beq
i\hbar \partial_t \rho = [H+L , \rho ] +
g_{ij}(K_i \vr K'_j \vr^\dagger - h.c. ).
\label{density}
\eeq
The ``primary'' object $\vr$ can not be eliminated from eq.
(\ref{density}) which therefore cannot be rephrased in terms of
$\rho$ only.
Thus unlike the case of eq. (1),
$\rho$ plays here the role of a ``secondary'' object.
Eq. (\ref{density}) also indicates that the term $L\vr$ in
eq. ({\ref{gLiouville}) gives rise to
a redefinition of the Hamiltonian and that the term $\vr R$ can be
eliminated. Indeed, the gauge transformation, $\vr \to \vr U$,
where $U$ is a unitary operator, does not affect expectation
values
and can be used to recast eq. (\ref{gLiouville}) into the form:
\beq
i\hbar \partial_t \vr = \tilde H \vr + g_{ij} K_i \vr \tilde K_j,
\label{gschr}
\eeq
where,  $\tilde H = H+L$,
$\tilde K_j = U K'_j U^{-1}$, and $U=\exp\biggl[-i\int^t (R-H)dt'
\biggr]$.
Without the last term this is simply a Schr\"odinger-like
equation
for the operator $\vr$.

To further analyze eq.  (\ref{gLiouville}) we construct  a Hilbert
space. It is defined as the linear space
$\l \equiv\lbrace \vr\rbrace$
of solutions of eq. (\ref{gLiouville}) with all possible
initial conditions at any $t_0$.  With the inner product defined
as:
\beq
\la\vr_1,\vr_2\ra =\tr \vr_1^\dagger\vr_2,
\label{inner}
\eeq
$\l$ becomes a Hilbert space.
It follows from eq. (\ref{gLiouville})
that this inner product is conserved and hence the
generalized dynamics suggested here manifests
in $\l$ as a unitary evolution.
The  inner product (\ref{inner}) may be regarded
as an extension of the ordinary quantum mechanical inner product.
If the corrections induced after $ t=t_0$  by the new terms in
the evolution eq. (\ref{gLiouville})
are small,
$\la\vr_1, \vr_2\ra\simeq |\la\psi_1|\psi_2\ra|^2$.
Note also that expression (\ref{prob}) for the expectation value of
of an observable $A$  can be now re-expressed  as:
\beq
\la A\ra  = {\la \vr, A\vr\ra \over \la\vr,\vr\ra}.
\label{exp}
\eeq

Equations (\ref{gLiouville},\ref{gschr}), and (\ref{inner}-\ref{exp})
suggest that $\vr$ should be interpreted as
a generalized ``wave operator''.
The new feature here however,
is that $\tr \rho^2=\tr(\vr\vr^\dagger)^2$ is not conserved.
This manifests the new aspects of our unitary evolution
as  transition between pure and mixed density matrices $(\rho)$.

The generalized unitarity, namely the conservation of the
inner product (\ref{inner}), can be clarified
by rewriting eq. (\ref{gLiouville}) in the Hilbert space $\l$.
For simplicity let us consider a system
with a finite, $N$-dimensional, Hilbert space and perform the
extension
described above.  The extended, $N^2$ dimensional, Hilbert space
$\l$ can be spanned by  a hermitian basis of $N^2-1$  SU(N)
matrices
and the unit operator:
\beq
\vr = {1\over\sqrt2} (\avr_0 {\bf 1} + \avr_i T_i),
\label{basis}
\eeq
where $T_i$ are SU(N) generators
and $\avr_a$ are $N^2$ complex numbers.
In this basis, the generalized inner product between any two
solutions is
given by an ordinary
vector product in a $N^2$-dimensional Hilbert space:
\beq
\la \vr_1, \vr_2\ra = \sum_{a=0}^{N^2-1}\avr_{1a}^* \avr_{2a}.
\eeq
We can also express eq. (\ref{gLiouville}) in this basis as a
Schr\"odinger-like equation:
\beq
i\hbar\partial_t \avr_a = \h_{ab} \avr_b
=(\h_{ab}^{(qm)}+\delta\h_{ab})\avr_b.
\label{gHam}
\eeq
The condition for conservation of probabilities (and unitarity)
is simply  that the  generalized Hamiltonian, $\h_{ab}$, is
hermitian. The deceptive similarity of eq. (\ref{gHam}) and
ordinary quantum
mechanics Schr\"odinger equation in  an $N^2$ dimensional space
notwithstanding,  we emphasis that the only relevant, physical
degrees of freedom are in those of the original ($N$ dimensional)
Hilbert space.

Next we would like to express the observables $A_i$ as
hermitian operators in $\l$.
In general we have in $\l$ \ \ $N^4$ independent hermitian
operators. Therefore  the mapping
\beq
A_i\to {\cal A}_i \in {\cal O}_\l
\label{mapping}
\eeq
of the original ($N^2$) observables $A_i$
 into the set of hermitian operators
${\cal O}_\l$ in  $\l$
is not  one to one. This mapping is constrained by demanding
that
\beq
\tr A_i\vr\vr^\dagger =
\sum_{a=0}^{N^2-1}\sum_{b=0}^{N^2-1}\avr_a ({\cal A}_i)_{ab}
\avr_b,
\label{map}
\eeq
i.e.,  that $\la A\ra$ is expressible in $\l$
as a ``standard''  expectation value
 with respect to the ``amplitudes'' $\vr_a$.
We also require that the mapping (\ref{mapping}) preserves  commutation
relations.
Therefore,
an $N$-dimensional
representation of SU(N) is mapped into an $N^2$ dimensional
representation of SU(N) in ${\cal O}_\l$, $T_i\to {\cal T}_i$.
The linear transformation maps a general  observable
$A_i=c_{i0} {\bf1} + c_{ia} T_a$ to
${\cal A}_i = c_{i0} {\cal I} + \sum_a c_{ia} {\cal T}_a$.
The operator
${\cal A}_i\in{\cal O}_\l$ still
has the same eigenvalues as the original operator $A_i$.
However all the eigenvalues
are now  $N$-fold degenerate. Another set of operators,
 denoted by
${\cal D}_j$, which remove the degeneracy of
${\cal A}_i$ do not correspond to observables.
It can be shown that the role of the new terms in
eq. (\ref{gLiouville}) or $\delta\h$ in eq. (\ref{gHam})
is to generate correlations between
${\cal A}_i$ and ${\cal D}_j$, which in turn induces the
transition to
a mixed density matrix.

It was noted by Gross \cite{Gross} and by Ellis et. al.
\cite{Ellis},
that linear modifications of the evolution laws for the density
matrix (e.g. eq.  (\ref{BPS})) generally
breaks  the one to one correspondence between
symmetries and conservation laws.
We now show
that in the present formalism, this correspondence is
partially restored.
An observable ${\cal A}\in {\cal O}_\l$ that is a
constant of motion satisfies:  $[{\cal A}, \h]=0$.
Hence the  unitary operator  $T=\exp(-i\epsilon {\cal A}/\hbar)$
commutes with the unitary evolution operator
$U=\exp({-it\h/\hbar})$,
and ${\cal A}$ generates a symmetry in $\l$.
The converse is not generally true. Since  $\l$
is $N^2$-dimensional, not all the hermitian operators in ${\cal
O}_\l$ may
be mapped back to hermitian operators in the original
$N$-dimensional Hilbert space.
Therefore, if some  hermitian operator ${\cal G}$ generates a
symmetry in $\l$ and its expectation value,
$\avr_{a}^*(t)\g_{ab}\avr_b(t)$,
is conserved, it still may not
correspond to an observable.

To illustrate the general discussion above let us
consider as an example the simple
two level system (e.g. a  spin half particle
in a constant magnetic field).
The mapping between the original 2-d Hilbert space and the 4-d
Hilbert space $\l$ will be spelled out in detail.
Let the ``free'' Hamiltonian be given by
\beq
H = E_0 + {1\over 2}\hbar\omega \sigma_3.
\label{free}
\eeq
We have seen that the terms  $L\vr$ and $\vr R$ in eq.
(\ref{gLiouville}) can be absorbed by a redefinition
of $H$ and $K'_j$. Therefore, the  modified eq. will be taken as:

\beq
i\hbar\partial_t \vr = [H, \vr] +   K\vr K',
\label{effham}
\eeq
where $K$ and $K'$ are functions of the Pauli matrices,
and will be assumed to be time independent.
Energy conservation,
${\partial\over\partial t}\la H\ra = {\partial\over\partial t}
{\la\vr, H\vr\ra\over\la\vr, \vr\ra}=0$, implies that
$[\sigma_3, K] = 0$, hence  $K= \sigma_3$.
This leaves three unknown parameters which determine $K'$:
\beq
K'=\a\sigma_1+\b\sigma_2+\lambda\sigma_3.
\eeq

When reexpressed in the four dimensional Hilbert space $\l$
the modified dynamics  corresponds to eq. (\ref{gHam}) with:
\beq
\d\h = \left( \matrix{
           \lambda   &  i\b  & -i\a   &     0    \cr
           -i\b  &  -\lambda  &     0       &  \a  \cr
           +i\a  &     0     &  -\lambda    &  \b  \cr
           0      &  \a  &  \b    &   \lambda  \cr
                     }\right) .
\label{spinmod}
\eeq

The observables in this model are combinations of $\sigma_i$ and
the
unit operator.
The mapping, $\sigma_i \to \s_i\in\o_\l$ is:
\beq
{1\over2}\sigma_k \to (\s_k)_{ab} =
       {1\over 2}(\d_{ak}\d_{b0} + \d_{a0}\d_{bk} +
i\epsilon_{abk}).
\label{smap}
\eeq
The $\s_i$ are a $4$-dimensional representation of $SU(2)$,
preserving the commutation relation
$[{\cal S}_i, {\cal S}_j]=i\epsilon_{ijk}{\cal S}_k$.
The mapping  (\ref{smap}) was constructed so as to
satisfy eq. ({\ref{map}).
The operators, ${\cal D}_j$, which remove the degeneracy of
${\cal S}_i$ have also been explicitly constructed.
The latter indeed do not correspond to observables.

It can now be verified that ${\cal S}_3$ is a constant
of  motion, i.e.  $[\s_3, \h^{qm}+\d\h]=0$.
We also notice that since the energy operator,
$E_0{\bf1}+ \hbar\omega\s_3$,
 is not the  mapped original  Hamiltonian:
$(\h^{qm})_{ab}=i\hbar\omega\epsilon_{ab3}$,
$\h^{qm}$ does not correspond to
an observable in $\l$.

The present model differs qualitatively from the model of
BPS or EHNS: while eq.
(\ref{BPS})  yields in general exponentially decaying (or
exponentially increasing)
solutions, our modifications  are oscillatory.
Indeed the general solution of equation (\ref{gHam}) is
\beq
\avr_a = \sum_{\mu=0,3} c_\mu\avr_{\mu a}
\exp(-i \lambda_{a} t),
\eeq
where $\lambda_\alpha$ and $\avr_{a\alpha}$  are the (real)
eigenvalues
and eigenvectors, respectively,  of $\h_{ab}$.

As an example consider the special case where only
$\lambda$ in eq. (\ref{spinmod}) is non-vanishing, and
the spin is  found at $t=t_0$ in the state $|\psi_0\ra
=\cos{\eta\over2}|\up_{\hat z} \ra + \sin{\eta\over2}|\down_{
\hat z} \ra$.
The solution in this case is given by
\beq
\vr(t) =
                \left( \matrix{
\cos^2(\eta/2) e^{-i\lambda t} & {1\over2}\sin(\eta)
e^{-i(\omega-\lambda)t} \cr
     {1\over2}\sin(\eta)  e^{i(\omega+\lambda)t}    &
\sin^2(\eta/2) e^{-i\lambda t}  \cr
                     }\right) .
\eeq
The resulting density matrix, $\rho=\vr\vr^\dagger$,
oscillates periodically between a pure and mixed state.
For example in the simple case $\eta=\pi/2$
\beq
\rho(t) = {1\over2}
                \left( \matrix{
           1         &   e^{-i\omega t}\cos(2\lambda t)   \cr
           e^{i\omega t}\cos(2\lambda t)      &  1         \cr
                     }\right),
\eeq
and $\tr\rho^2 = {1\over2}+{1\over2}\cos^2(2\lambda t)$.

Observable effects due to these modifications
can in principle be searched for
in  neutron interferometry experiments \cite{neutron}.
In such interference  experiments,
one typically measures an observable of the form
\beq
A(\theta) = {1\over2}
                \left( \matrix{
           1         &   e^{i\theta}   \cr
           e^{-i\theta}      &  1         \cr
                     }\right) ,
\eeq
where $\theta$ is determined by the experimental set up.
The expectation value of $A$ is given in our case by
\beq
\la A \ra = {1\over2}\biggl[1 + \sin\eta\Bigl[
\cos^2(\eta/2)\cos((\omega+2\lambda) t+\theta) +
 \sin^2(\eta/2)\cos((\omega-2\lambda) t+\theta)\Bigr] \biggr].
\label{average}
\eeq
The correction is indeed oscillatory.
This should be contrasted with the exponential
$\exp(-2\lambda_{ENHS} t)$
decay of the interference obtained by EHNS.

What are the present experimental bounds pertinent to the three
new
parameters of the two level system?
We can use the two slit experiments
 of Zeilinger et. al. \cite{exp}
with a $20 A^0$ neutron beam, and the analysis of Pearle
\cite{Pearle},
 to constrains the generic
parameter $\lambda$ to $\lambda \sim 10 (sec)^{-1} \sim 10^{-23}$
GeV.
The constrains  of the same experiment on the corresponding
parameters
in the EHNS model is $\sim 100$ times stronger ($\sim 10^{-25}$
GeV).
The exponential factor modifies the interference contrast during
the short flight time  ($t_0\simeq 10^{-2}$ sec)
by $(1-2\lambda_{EHNS}t_0)$. In the present case the extra
oscillation can
be subsumed into slow ``beating'' $\sim \cos(2\lambda t_0) \simeq
(1-2(\lambda t_0)^2)$,  causing a much weaker reduction
of the contrast in the interference pattern.

We found that
our  modification induces $K_L$ $K_S$ mixing
generating CP violation
in the two level $K_0-\bar K_0$ system in a  similar
fashion as in the EHNS model.
However, this mixing predicts a phase
of the  CP violating parameter $\epsilon$,
which is of by $\pi/2$ just as in the case of
the EHNS model \cite{Huet}.
Hence our modification can account only for a small part
of the CP violation observed in the $K_0-\bar K_0$ system.
This leads to the generic upper bound of order
$\sim 10^{-19}$ Gev, of the same order as
$M_K^2/M_{pl}$  which could be expected on dimensional
grounds if CP and/or CPT violations are due to
effects of quantum gravity.
The hundred fold larger parameter allowed by the neutron
interference
experiment in our model could be important.
In particular this renders smaller yet  experimentally detectable
CPT violations,  more likely in the present framework.

We have constructed a formalism based on an operator
generalization  of the wave function which is linear, local, and unitary.
As a consistency check of this proposal we note that to  some
extent the proposed formalism can be embedded in the framework of
ordinary quantum mechanics.
We can interpret $\vr\vr^\dagger$ and $\vr^\dagger\vr$
as the reduced density matrices of a sub-system
and an environment, respectively.
The generalized Hamiltonian $\h_{ab}$ in eq. (\ref{gHam}) and the
amplitudes $\avr_a$ can then be interpreted
as the Hamiltonian  and wave function of the  total system,
while the  new terms in eq. (\ref{gLiouville},\ref{gschr}) or (\ref{gHam})
as describing an  interaction between the sub-system and the environment.
Therefore, the consistency of the proposed equation of motion
follows from  quantum mechanics.
{\it Nevertheless}, postulate  (4), $\vr(t=t_0) = |\psi_0\ra\la\psi_0|$,
 which sets the initial condition for eq. (\ref{gLiouville})
goes beyond any ordinary quantum mechanical scheme.
It would amount in quantum mechanics to an  additional requirement that
after carrying a complete set of measurements on the sub-system,
the wave function of the environment becomes identical
to that of the system. This additional
constraint is not satisfied in quantum mechanics.
Therefore the predictions of this formalism will
generally differ from that of a quantum mechanical system
with an  environment \cite{note3}.

Finally, we note that  the proposed formalism may also be relevant to the
information problem  in black hole evaporation and to the measurement
problem. In the latter case, for large systems the
modified evolution  might under appropriate conditions give rise to
loss of coherence which amounts to a measurement.

\vspace{1in}
{\bf Acknowledgement}

I have benefited from discussions and helpful
comments of Y. Aharonov, A. Casher, S. Coleman, P. Huet,
S. Nussinov, S. Popescu, W. G. Unruh, and N. Weiss.

\vfill\eject


\begin{thebibliography}{99}
\bibitem{note}
This procedure is not uniquely defined
since we could have postulated the initial condition
$\vr(t=t_0) = |\psi\ra\la u|$,
where $|u\ra$ is some arbitrary fixed state.
\bibitem{note2}
It is interesting to note that (non-linear) odd power of $\vr$
can still be added to eq. (7),
which still allow conservation of  probability.
For example, the most general cubic term with such property
has the form $A\vr B\vr^\dagger B \vr C$
where $A$, $B$, $C$ are hermitian operators. However such terms
do no longer conserve the inner product eq. (10)
and shall not be discussed further.
\bibitem{note3}
Although the oscillatory corrections found in the two-level
system are common also
in the case of a coupling with some environment, the two cases
are distinguishable. The modified equation of motion of the
two-level model under consideration  can be mapped to
an ordinary quantum mechanics spin half sub-system which is
coupled to another spin half (the environment).
The interaction term is then given by
$H_{int} = \lambda \sigma_z^1\sigma_z^2$.
Assuming that a-priori no information about the initial
state of $\sigma^2$ is known,  we find that quantum mechanics
yields a prediction which is identical to $\la A \ra $ in  eq. (26)
only in the special  cases of $\eta=n{\pi\over2}$.
\end{thebibliography}
\end{document}